\documentclass[aps,prd,preprint,floatfix,nofootinbib]{revtex4}

\usepackage{graphicx,epsfig}

\newcommand{\beq}{\begin{equation}}
\newcommand{\eeq}{\end{equation}}
\newcommand{\bea}{\begin{eqnarray}}
\newcommand{\eea}{\end{eqnarray}}
\newcommand{\nn}{\nonumber}
\newcommand{\ie}{{\it i.e.}}

\def\ponedp{\Phi_1^\dagger \Phi_1}
\def\ptwodp{\Phi_2^\dagger \Phi_2}
\def\pidp{\Phi_i^\dagger \Phi_i}
\def\pjdp{\Phi_j^\dagger \Phi_j}
\def\ede{\eta^\dagger \eta}
\def\rmre{{\rm Re}}
\def\rmim{{\rm Im}}
\def\rmSM{{\rm SM}}
\def\sbet{s_\beta}
\def\cbet{c_\beta}
\def\sE{s_E}
\def\cE{c_E}

\def\esq{\epsilon^2}

\begin{document}

\begin{flushright}
OITS-780\\
%hep-ph/yymmnnn\\
\end{flushright}

\bibliographystyle{revtex}

\title{The Family $SU(2)_l \times SU(2)_h\times U(1)$ Model}

\author{Cheng-Wei Chiang$^{1,2}$, N.G. Deshpande$^3$,Xiao-Gang He$^4$,  J. Jiang$^3$}
\affiliation{$^1$Department of Physics and Center for Mathematics and Theoretical Physics, National Central University, Chungli, Taiwan 320, ROC}
\affiliation{$^2$Institute of Physics, Academia Sinica, Taipei, Taiwan
115, ROC}
\affiliation{$^3$Institute for Theoretical Science, University of Oregon,
Eugene, OR 97403, USA}
\affiliation{$^4$Department of Physics and Center for Theoretical Sciences, National Taiwan University, Taipei, Taiwan, ROC}\bigskip
\date{\today}

\begin{abstract}
We consider extension of the standard model $SU(2)_l \times SU(2)_h \times U(1)$ where the first two families of quarks and leptons transform according to the $SU(2)_l$ group and the third family according to the $SU(2)_h$ group.  In this approach, the largeness of top-quark mass is associated with the large vacuum expectation value of the corresponding Higgs field.  The model predicts almost degenerate heavy $W'$ and $Z'$ bosons with non-universal couplings, and extra Higgs bosons.  We present in detail the symmetry breaking mechanism, and carry out the subsequent phenomenology of the gauge sector.  We compare the model with electroweak precision data, and conclude that the extra gauge bosons and the Higgs bosons whose masses lie in the TeV range, can be discovered at the LHC.
\end{abstract}

\maketitle

\section{Introduction}

As we enter the era of the Large Hadron Collider (LHC), we anticipate the discovery of new physics (NP).  In the past decade, we have witnessed many interesting theoretical proposals, each with its own variety of new particles beyond the standard model (SM).  Several of these proposals require extra gauge bosons, for example, from a larger gauge group \cite{lgmodels}, from extension to higher dimensions \cite{ldim} which leads to Kaluza-Klein type of mass ladders,
%CC added following phase
or from non-commuting extended technicolor \cite{NCETC}.  Extensions of SM with additional $W$'s and $Z$'s that have non-universal couplings to quarks and leptons have also been considered.  In this paper, we analyze a model with extra weak gauge bosons from the consideration of family structure.

The electroweak (EW) gauge group of our model is $SU(2)_l \times SU(2)_h \times U(1)_Y$, where $l$ and $h$ stand for light and heavy families, respectively.  The first two quark and lepton families are considered as light while the third as heavy.  For each $SU(2)$ gauge group, the chiral fermionic particles are the same as the SM particle contents and, therefore, the model is anomaly-free.  In this framework, the large mass of the top quark is induced by a large vacuum expectation value (VEV) of one Higgs field responsible for $SU(2)_h$ breaking.  A logical extension of the idea would have been to consider one $SU(2)$ for each family. Such an idea has already been proposed some time back by Li and Ma where $SU(2)$ for each generation was introduced \cite{ma}. With appropriate symmetry breaking patterns, the $SU(2)_l \times SU(2)_h \times U(1)_Y$ model can be produced.  Later several authors have considered the same model and studied some consequences of this model \cite{NCETC,TFmodels,EWP}.
%CC added following sentences
Some low energy phenomenological~\cite{TFpheno} and cosmological~\cite{Bviolation} consequences have also been analyzed.

The mechanism of generating the mass for the top and the Higgs structure in the above-mentioned papers differ from our treatment here.  The mechanism in the $SU(2)_l \times SU(2)_h \times U(1)_Y$ model that we are considering is a more conventional approach with an explicit Higgs structure. We shall first carry out the consequences of the breaking of symmetry, then study the Yukawa, gauge interactions and FCNC interactions in these sectors, and finally analyze the phenomenological consequences.  Our study of the Higgs structure clarifies conditions necessary for the light Higgs to be flavor conserving.  We also impose tight constraints based on electroweak precision (EWP) data, where standard model radiative corrections along with new physics to the lowest order perturbatively are included. The allowed masses of gauge bosons and Higgs are far more restricted as a consequence.

We start with the EW group of $SU(2)_1 \times SU(2)_2\times U(1)_Y$ at a high-energy scale of the order of a few TeV.  For ease of notation, we hereafter use indices 1 and 2 for $l$ and $h$, respectively.  The first two families are charged under $SU(2)_1$, and the third family is charged under $SU(2)_2$.  We note that such a group structure can arise from a broken grand unified model based on $SU(3)^3$ or $SU(15)$.  We do not pursue this issue here though.  The quarks, leptons and Higgs bosons and their gauge group representations in our model are as follows:
\begin{eqnarray}
&&Q_{jL}: (2,1)(1/3)\;,\;\;Q_{3L}: (1,2)(1/3)\;,\;\;U_{iR}: (1,1)(4/3)\;,\;\;D_{iR}:(1,1,)(-2/3)\;,\nonumber\\
&&L_{iL}: (2,1)(-1)\;,\;\;L_{3L}: (1,2)(-1)\;,\;\;E_{iR}: (1,1)(-2)\;,\\
&&\Phi_1:(2,1)(1)\;,\;\;\Phi_2: (1,2)(1)\;,\;\;\eta: (2,2)\;,\nonumber
\end{eqnarray}
where the two numbers in the first parentheses indicate the $SU(2)_1$ and $SU(2)_2$ representations, respectively, and the number in the second parentheses gives the $U(1)_Y$ quantum number.

We require that the gauge group is broken to the SM gauge group of $SU(2)_L \times U(1)_Y$ first, and then further broken to the $U(1)_{\rm EM}$ group.  These are realized by the non-zero VEV of the Higgs fields.  The self-dual bi-linear Higgs field $\eta$, charged under both $SU(2)$ gauge groups, acquires a VEV, $\langle \eta \rangle = {\rm diag(u, u)}$, at scale $u$ and breaks the $SU(2)_1 \times SU(2)_2$ group to the diagonal $SU(2)_L$.  The gauge bosons corresponding to the broken generators develop masses of order $u$.  The other gauge bosons and fermions remain massless at this point.  The coupling of the surviving $SU(2)_L$ is $g$, with
\beq
\frac{1}{g^2} = \frac{1}{g_1^2} + \frac{1}{g_2^2}~.
\eeq

The next stage of symmetry breaking is achieved by the non-zero VEV's $v_i$ of $\Phi_i$, breaking the remaining $SU(2)\times U(1)_Y$ to the $U(1)_{\rm EM}$ and rendering the usual $W$ and $Z$ bosons and nonzero fermion masses.  The coupling of the surviving $U(1)_{\rm EM}$ is $e$, with
\beq
\frac{1}{e^2} = \frac{1}{g^2} + \frac{1}{g^{\prime 2}} = \frac{1}{g_1^2} + \frac{1}{g_2^2}+ \frac{1}{g^{\prime 2}}\;,
\eeq
where $g^\prime$ is the coupling of the $U(1)_Y$ gauge group.

The Weinberg angle $\theta_W$ is defined by $x_0 =\sin^2\theta_W = g^{\prime 2}/(g^2+g^{\prime 2})$.  We will use $s_W$ and $c_W$ for the sine and cosine of $\theta_W$, respectively.  For convenience, we also define a mixing angle of the extended gauge group, $\theta_E$, with sine ($s_E$) and cosine ($c_E$) of this angle given by $c_E= g/g_1$ and $s_E=g/g_2$.  For the VEV's of the doublets, we define an angle $\beta$ with $\tan\beta = v_2/v_1$.

The structure of this paper is as follows.  In Section~\ref{sec:higgs}, we present a detailed analysis of the Higgs potential and the Higgs mass spectrum.  Following that, we give the Yukawa couplings of the fermions and their mixing in Section~\ref{sec:yukawa}.  In Section~\ref{sec:gauge}, we compute the gauge boson mass spectrum and their interactions with fermions at tree level.  We then analyze the phenomenological constraints from EWP data, lepton universality, atomic parity violation, and flavor-changing neutral currents (FCNC's) in Section~\ref{sec:pheno}.  We summarize our findings in Section~\ref{sec:summary}

\section{The Higgs Potential and the Higgs Boson Masses
\label{sec:higgs}}

In this section, we provide some ideas about the Higgs boson masses in the model.  The most general Higgs potential is given by
\bea
V = && \sum \mu_i^2 \pidp + \frac{1}{4} \sum \lambda_{ij}
(\pidp)(\pjdp) + M^2 Tr(\ede) + Tr(\tilde M^2 \tilde \eta \eta + h.c.)\nonumber\\
\label{eq:HiggsV}
&& + \frac{1}{4} h [Tr(\ede)]^2 + \frac{1}{4} (\tilde h [Tr(\tilde \eta \eta)]^2 + h.c.) + Tr(\ede)Tr(\tilde f \tilde \eta \eta + h.c.)\\
&& + \frac{1}{2} \sum f_i (\pidp) Tr(\ede) + \sum p_i \Phi_i^\dagger \eta
\eta^\dagger \Phi_i + \sum (\tilde p_i \Phi_i^\dagger \tilde \eta \eta \Phi_i + h.c.)\nonumber\\
&&+( t' \Phi_1^\dagger \eta \Phi_2 + h.c) + (\tilde t \Phi_1^\dagger \tilde \eta \Phi_2 + h.c)\;,\nonumber
\eea
where $\tilde \eta = \sigma_2 \eta^* \sigma_2$ and $\sigma_2$ is one Pauli matrix. If no CP violation originates from the Higgs potential, all the coefficients will be real, as we will assume in our latter discussions.

One can carry out a full detailed analysis for the Higgs mass spectrum with the above complete potential. Here we will provide a simplified analysis by noticing that the VEV $u$ is much larger than the VEVs $v_i$ and that the fields in $\eta$ become heavy and almost decouple from the fields in $\Phi_i$.  The fields that couple to fermions and therefore have possible large observable effects are the $\Phi_i$ fields.  We can approximate the Higgs potential involving $\Phi_i$ by replacing $\eta$ with its VEV $u$ in Eq.~(\ref{eq:HiggsV}).  The effective Higgs potential is now
\bea
V = && m_1^2 \ponedp + m_2^2 \ptwodp + \frac{1}{4}\lambda_1 (\ponedp)^2 +
\frac{1}{4} \lambda_2 (\ptwodp)^2 \nn \\
&& + \frac{1}{2} \lambda_{12} (\ponedp)(\ptwodp) +  t u (\Phi_1^\dagger \Phi_2 +
\Phi_2^\dagger \Phi_1)~,
\eea
where $m_1^2 = \mu_1^2 + (f_1+p_1+\tilde p_1) u^2$, $m_2^2 = \mu_2^2 +( f_2+p_2+\tilde p_2) u^2$, $\lambda_1 = \lambda_{11}$, $\lambda_2 = \lambda_{22}$, and $t = t' + \tilde t$.

We now proceed to to the next stage when $\Phi_1$ and $\Phi_2$ acquire VEV's $v_1$ and $v_2$
\beq
\langle \Phi_1 \rangle = \left( \begin{array}{c}
	0 \\
	v_1
\end{array} \right)~, \quad {\rm and} \quad
\langle \Phi_2 \rangle = \left( \begin{array}{c}
	0 \\
	v_2 e^{i \xi}	
\end{array} \right)~,
\eeq
where $v_1^2 + v_2^2 = v^2$ with $v$ being the VEV related to electroweak symmetry breaking close to 174 GeV in the SM.  Expanding $\Phi_1$ and $\Phi_2$ with respect to their VEV's
\beq
\Phi_1 = \left( \begin{array}{c}
	\phi_1^+ \\
	v_1 + {\rm Re}\phi_1^0 + i {\rm Im}\phi_1^0
\end{array} \right)~, \quad {\rm and} \quad
\Phi_2 = \left( \begin{array}{c}
	\phi_2^+ \\
	v_2 e^{i \xi} + {\rm Re}\phi_2^0 + i {\rm Im}\phi_2^0	
\end{array} \right) ~,
\eeq
the Higgs potential now becomes
\bea
V = && m_1^2\,[(\phi_1^+)^2 + (v_1 + \rmre \phi_1^0)^2 + (\rmim
\phi_1^0)^2] + m_2^2[(\phi_2^+)^2 + ( v_2 \cos\xi + \rmre \phi_1^0)^2
\nn \\
&& \quad \quad + (v_2
\sin\xi + \rmim \phi_2^0)^2] \nn \\
&& + \frac{1}{4} \lambda_1 [(\phi_1^+)^2 + (v_1 + \rmre \phi_1^0)^2 + (\rmim
\phi_1^0)^2]^2 + \frac{1}{4} \lambda_2 [(\phi_2^+)^2 \nn \\
&& \quad \quad + (v_2 \cos\xi + \rmre
\phi_2^0)^2 + (v_2
\sin\xi + \rmim \phi_1^0)^2]^2  \\
&& + \frac{1}{2} \lambda_{12} [(\phi_1^+)^2 + (v_1 + \rmre \phi_1^0)^2 + (\rmim
\phi_1^0)^2][(\phi_2^+)^2 + (v_2 \cos\xi + \rmre \phi_2^0)^2 \nn \\
&& \quad \quad + (v_2
\sin\xi + \rmim \phi_1^0)^2] \nn \\
&& + t u \,[(\phi_1^+)^* \phi_2^+ + (v_1 + \rmre \phi_1^0 - i\, \rmim
\phi_1^0) ( v_2 e^{i \xi} + \rmre \phi_2^0 + i\, \rmim \phi_2^0) \nn \\
&& + (\phi_2^+)^* \phi_1^+ + ( v_2 e^{- i \xi} + \rmre \phi_2^0 - i\,
\rmim \phi_2^0) (v_1 + \rmre \phi_1^0 + i\, \rmim \phi_1^0) ]~.\nn
\eea
In the above expression, we have removed a constant term proportional to powers of the VEV of $\eta$ and terms associated with $\eta$ field fluctuating around the VEV.

The stability condition requires that $\sin\xi = 0$.  The sign of $\cos\xi$ depends on the sign of $t$, $t \cos\xi = - |t|$.  The stability conditions on $v_1$ and $v_2$ are
\bea
&& 2 m_1^2 v_1 + \lambda_1 v_1^3 + \lambda_{12} v_1 v_2^2 - 2 |t| u v_2
= 0~, \nn \\
&& 2 m_2^2 v_2 + \lambda_2 v_2^3 + \lambda_{12} v_1^2 v_2 - 2 |t| u v_1
= 0~.
\eea
Hence, the mass-squared matrices of $\phi_{1,2}^+$ and $\rmim \phi_{1,2}^0$ turn out to be identical and are
\begin{eqnarray}
M_{\phi^+}^2 &=& M_{\rmim \phi^0}^2 = \left(
\begin{array}{cc}
m_1^2 + \frac{1}{2} \lambda_1 v_1^2 + \frac{1}{2} \lambda_{12} v_2^2 &
t u \\
t u & m_2^2 + \frac{1}{2} \lambda_2 v_2^2 + \frac{1}{2} \lambda_{12}
v_1^1
\end{array}
\right)\nn\\
&=&
\left( \begin{array}{cc}
\frac{v_2}{v_1} |t| u & t u \\
t u & \frac{v_1}{v_2} |t| u
\end{array}
\right)~.
\end{eqnarray}
There are massless Goldstone modes associated with both $\phi^+$ and $\rmim \phi^0$.  At the tree level, $\phi^{\pm}$ and $A^0$ have the same mass
\beq
m_{\phi^{\pm}}^2 = m_{A^0}^2 = \frac{v^2}{v_1 v_2} |t| u ~.
\eeq

The mass-squared matrix for neutral Higgs bosons is
\bea
M_{\rmre \phi^0}^2 = && \left(
\begin{array}{cc}
m_1^2 + \frac{3}{2} \lambda_1 v_1^2 + \frac{1}{2} \lambda_{12} v_2^2 &
\frac{1}{2} \lambda_{12} v_1 v_2 + t u\\
\frac{1}{2} \lambda_{12} v_1 v_2 + t u  & m_2^2 + \frac{3}{2}
\lambda_2 v_2^2 + \frac{1}{2} \lambda_{12} v_1^2
\end{array}
\right) \nn \\
= &&
\left( \begin{array}{cc}
\frac{v_2}{v_1} |t| u  + \lambda_1 v_1^2 & \lambda_{12} v_1 v_2 + t u \\
\lambda_{12} v_1 v_2 + t u & \frac{v_1}{v_2} |t| u + \lambda_2 v_2^2
\end{array}
\right)~.
\eea
In the two Higgs doublet models, there generally exist flavor-changing neutral currents (FCNC's) when both doublets acquire VEV's.  To better understand the FCNC structure, it is convenient to work in the basis where the Goldstone bosons are singled out by the following rotation,
\beq
\label{eq:higgsrot}
\left(
\begin{array}{c}
\Psi_1 \\
\Psi_2
\end{array}
\right) =
\left(
\begin{array}{cc}
\cbet & \sbet \\
-\sbet & \cbet
\end{array}
\right)
\left(
\begin{array}{c}
\Phi_1 \\
\Phi_2
\end{array}
\right)~.
\eeq
Now only $\Psi_1$ acquires a VEV $v$,
%
%\beq \langle \Psi_1 \rangle = \left( \begin{array}{c}
%	0 \\
%	v
%\end{array} \right)~, \quad {\rm and} \quad
%\langle \Psi_2 \rangle = \left( \begin{array}{c}
%	0 \\
%	0	
%\end{array} \right)~.
%\eeq
%
Expansions of $\Psi_1$ and $\Psi_2$ around their VEV's are
\beq
\label{eq:higgscomp}
\Psi_1 = \left( \begin{array}{c}
	G^+ \\
	v + h + i G^0
\end{array} \right)~, \quad {\rm and} \quad
\Psi_2 = \left( \begin{array}{c}
	H^+ \\
	H^0 + i A^0	
\end{array} \right)~,
\eeq
where $G^+$ and $G^0$ are the Goldstone bosons, $H^+$ the charged Higgs boson, $A^0$ the pseudoscalar boson, and $h$ and $H^0$ the neutral light and heavy scalar bosons, respectively.  Note that in the reduced effective potential, $G^+$ and $G^0$ correspond to the Goldstone bosons ``eaten'' by the $W$ and $Z$ bosons. In the full theory, there will in general be mixings with component fields in $\eta$.  The physical Higgs mass-squared matrices are
\bea
&& M_{H^+}^2 = M_{A^0}^2 = \left(
\begin{array}{cc}
0 & 0 \\
0 & \frac{1}{\sbet \cbet } |t| u
\end{array}
\right)~, \nn \\
&& M_{h, H}^2 = v^2 \sbet \cbet  \\
&& \left(
\begin{array}{cc}
  \frac{1}{\sbet\cbet} (\lambda_1 \cbet^4 + \lambda_{12} \sbet^2
  \cbet^2 + \lambda_2 \sbet^4)
 & -\lambda_1 \cbet^2 -  \lambda_{12} \sbet^2  +
  \lambda_{12} \cbet^2 + \lambda_2 \sbet^2  \\
  - \lambda_1 \cbet^2 -  \lambda_{12} \sbet^2  +
  \lambda_{12} \cbet^2 + \lambda_2 \sbet^2
 & -  \frac{1}{\sbet^2 \cbet^2} \frac{t u}{v^2} +  \sbet \cbet
  (\lambda_1 - 2 \lambda_{12} + \lambda_2 )
\end{array}
\right)~.\nn
\eea
To reduce FCNC's mediated by the SM Higgs boson, we need to suppress the $h$-$H$ mixing since $H$ will induce tree-level FCNC interactions in the Yukawa couplings.  To ensure that the off-diagonal terms vanish, it is required that $-\lambda_1 \cbet^2 - \lambda_{12} \sbet^2 + \lambda_{12} \cbet^2 + \lambda_2 \sbet^2 = 0$, or at least be very small.  In a specific realization of this condition, $\lambda_1 = \lambda_2 = \lambda_{12} = \lambda$, the mass-squared matrix for the $\rmre \phi^0$ fields is
\beq
M_{h,H}^2 =
\left(
\begin{array}{cc}
2 \lambda v^2 & 0 \\
0 & - \frac{t u}{\sbet \cbet}
\end{array}
\right)~.
\eeq
We therefore have a SM-like Higgs field $h$, and a degenerate heavy scalar doublet whose mass can be in the TeV range. Since the heavy Higgs can mediate flavor changing processes, we will address mass constraints on this field in Section \ref{sec:pheno}-B.

\section{Yukawa Interactions\label{sec:yukawa}}

The Yukawa interactions are
\bea
{\mathcal L}_{\rm Yukawa} = && f_{ij}^u \bar{u}_{iR}
{\tilde{\Phi}_1}^\dagger Q_{jL}
+ g_{i3}^u \bar{u}_{iR}
{\tilde{\Phi}_2}^\dagger Q_{3L}
+ f_{ij}^d \bar{d}_{iR}
{\Phi}_1^\dagger Q_{jL}
+ g_{i3}^d \bar{D}_i
{\Phi}_2^\dagger Q_{3L}~,
\label{eq:Yukawa}
\eea
where the family index $i$ sums over $1, 2, 3$ and $j$ sums over $1, 2$, the field $u_{iR}$ denotes right-handed up-type quarks and $d_{iR}$ the right-handed down-type quarks, and $Q_{jL} = (u_{jL}, d_{jL})^T$ and $Q_{3L} = (u_{3L}, d_{3L})^T$ are left-handed quark doublets.  Here $f_{ij}$ and $g_{ij}$ are the Yukawa couplings, and $\tilde{\Phi}$ is defined as $\tilde{\Phi} = i \sigma_2 \Phi$.  Substituting Eqs. (\ref{eq:higgsrot}) and (\ref{eq:higgscomp}) into Eq.~(\ref{eq:Yukawa}), we have
\bea
{\mathcal L}_{\rm Yukawa} &=& - \bar{U}_{R} M^u U_L (1 + \frac{h}{v})
- \bar{D}_R M^d D_L (1 + \frac{h}{v})\nn \\
&& + \bar U_R (\lambda^u_1 - \lambda^u_2) U_L (H^0 - iA^0) + \bar D_R (\lambda^d_1 - \lambda^d_2) D_L(H^0 + i A^0)\\
&& - \bar U_R (\lambda^u_1 - \lambda^u_2) D_L H^+ + \bar D_R (\lambda^d_1 - \lambda^d_2)U_L H^- + h.c.~,\nn
\eea
where $U^T_{L,R} = (u,\;c,\;t)_{L,R}$ and $D^T_{L,R} = (d,\;s,\;b)_{L,R}$. The coupling matrices $\lambda_i^{u,d}$ and the mass matrices $M^{u,d}_i$
are given by
\beq
\lambda_1^u = \left(
\begin{array}{ccc}
f_{11}^u &  f_{12}^u & 0 \\
f_{21}^u &  f_{22}^u & 0 \\
f_{31}^u &  f_{32}^u & 0
\end{array}
\right)~,\;\;
\lambda_2^u = \left(
\begin{array}{ccc}
0 & 0 &  g_{13}^u \\
0 & 0 &  g_{23}^u \\
0 & 0 &  g_{33}^u
\end{array}
\right)~,\;\; M^u = v(\cbet \lambda_1^u + \sbet \lambda^u_2) ~,
\eeq
and
\beq
\lambda^d_1  = - \left(
\begin{array}{ccc}
 f_{11}^d &  f_{12}^d & 0 \\
 f_{21}^d &  f_{22}^d & 0 \\
 f_{31}^d &  f_{32}^d & 0
\end{array}
\right)~,\;\; \lambda^d_2  = - \left(
\begin{array}{ccc}
0 & 0 & g_{13}^d \\
0 & 0 & g_{23}^d \\
0 & 0 & g_{33}^d
\end{array}
\right)~,\;\;M^d = v(\cbet \lambda_1^d + \sbet \lambda^d_2)\;.
\eeq

It is clear that if $v_2$ is much larger than $v_1$, one can naturally explain why the third-generation quark masses are much larger than those in the first two generations.

The quark mass matrices can be diagonalized by bi-unitary transformations of the following form
\beq
S_U^\dagger M^u T_U = {\rm diag}\{m_u, m_c, m_t\} = \hat M^u~,
\quad
{\rm and}
\quad
\label{eq:CKM}
S_D^\dagger M^d T_D = {\rm diag}\{m_d, m_s, m_b\} = \hat M^d~.
\eeq
In the quark mass eigenstate basis, we have
\bea
{\mathcal L}_{\rm Yukawa} &=& - \bar{U}_{R} \hat M^u U_L (1 + \frac{h}{v})
- \bar{D}_R \hat M^d D_L (1 + \frac{h}{v})\nn \\
&& + \bar U_R \lambda^u U_L (H^0 - iA^0) + \bar D_R \lambda^d D_L(H^0 + i A^0)\\
&& - \bar U_R \lambda^u V_{KM} D_L H^+ + \bar D_R \lambda^d V_{KM}^\dagger U_L H^- + h.c.~,\nn
\eea
where $\lambda^u = S_U(\lambda^u_1 - \lambda^u_2)T^\dagger_U = -M^u/vs_\beta + (1+c_\beta/s_\beta)S_U \lambda^u_1 T^\dagger_U$
and $\lambda^d = S_D(\lambda^d_1 - \lambda^d_2)T^\dagger_D = -M^d/vs_\beta + (1+c_\beta/s_\beta)S_D \lambda^d_1 T^\dagger_D$.  Here
$V_{KM} = T_U T_D^\dagger$ is the Cabbibo-Kobayashi-Maskawa (CKM) mixing matrix.

It is not possible to solve for these matrices of the model without specifying $f_{ij}$ and $g_{ij}$.  For some simplified cases, one can completely know the FCNC structure by Higgs exchange, for example: a) $S_U = T_U = S_D = 1$, then $T_D = V^\dagger_{KM}$, and b) $S_D = T_D = S_U = 1$, then $T_U = V_{KM}$.  In case a), $M^u = \hat M^u V_{KM}$ and in case b), $M^d = \hat M^d V_{KM}^\dagger$. The coupling matrices in these two cases are then completely determined by the quark eigen-masses and the CKM matrix.

 One can also easily work out the couplings in the lepton sector. The results are similar to the quark sector and can be obtained by replacing $D_{L,R}$ with $E_{L,R} = (e_{L,R},\;\mu_{L,R,}\;\tau_{L,R})$. If three right-handed neutrinos $\nu_R = (\nu_{R1}, \nu_{R2},\;\nu_{R3})^T$ are introduced into the theory, then the relevant Yukawa couplings can be obtained by replacing $U_{L,R}$ by $\nu_{L,R}
= (\nu^e_{L,R},\;\nu^\mu_{L,R},\;\nu^\tau_{L,R})^T$.

Note that the tree-level FCNC's are associated with the heavy Higgs bosons, $H^0$ and $A^0$, and the Yukawa couplings are given by $(1+c_\beta/s_\beta)S_i \lambda^i_1 T^\dagger_i$. We will comment on the constraints from FCNC data on the Higgs masses and Yukawa couplings when we study the phenomenology in Section~\ref{sec:pheno}.

\section{Gauge Interactions \label{sec:gauge}}

Gauge bosons interact with Higgs and fermions through the covariant derivative terms:
\begin{eqnarray}
(D_\mu \Phi_i)^\dagger (D^\mu \Phi_i)\;,\;\; Tr[(D_\mu \eta)^\dagger (D^\mu \eta)]\;,\;\; i \bar \psi \gamma_\mu D^\mu \psi\;,
\end{eqnarray}
where $\psi$ indicates a generic fermion fields in the model. The covariant derivatives are given by
\begin{eqnarray}
&&i D^\mu \phi_i = (i \partial^\mu + {g_1\over 2} W_1^\mu + {g_2\over 2} W_2^\mu + {g^\prime\over 2} Y B^\mu) \Phi_i\;,\nonumber\\
&&i D^\mu \psi = (i \partial^\mu + {g_1\over 2} W_1^\mu + {g_2\over 2} W_2^\mu + {g^\prime\over 2} Y B^\mu) \psi\;,\\
&&i D^\mu \eta = (i \partial^\mu - {g_1\over 2} W_1^\mu + {g_2\over 2} W_2^\mu) \eta\;,\nn
\end{eqnarray}
where $W^\mu_i = W^{\mu a}_i \sigma_a$ with $\sigma_a$ the Pauli matrices.

After the Higgs boson fields develop VEV's, the gauge bosons corresponding to the broken generators will become massive.  We obtain the mass-squared matrix for the charged gauge bosons in the $(W_1, W_2)$ basis as follows:
\beq
M_W^2 = \frac{1}{2} \left(
\begin{array}{cc}
g_1^2 (v_1^2 + 2 u^2) & -2 g_1 g_2 u^2 \\
-2 g_1 g_2 u^2 & g_2^2 (v_2^2 + 2 u^2)
\end{array}
\right)~.
\eeq

Since the large VEV $u$ breaks the $SU(2)_1 \times SU(2)_2$ to a diagonal $SU(2)_L$, it is convenient to work in a basis $(W_H, W_L)$.  In the limit that $v_i$ go to zero, the mass of $W_L$ goes to zero and it can be identified as one of the gauge boson fields in the unbroken $SU(2)_L$.  The relations between $W_{1,2}$ and $W_{L,H}$ are
\beq
W_1 = \frac{g_2 W_L + g_1 W_H}{\sqrt{g_1^2 + g_2^2}},\quad
{\rm and} \quad
W_2 = \frac{g_1 W_L - g_2 W_H}{\sqrt{g_1^2 + g_2^2}}~,
\eeq
The $W_{L,H}$ mass-squared matrix, with non-zero $v_i$, becomes
\beq
M_W^2 =
\frac{1}{2} \left(
\begin{array}{cc}
 (g_1^2 + g_2^2) u^2 + \frac{g_1^4 v_1^2 + g_2^4 v_2^2}{g_1^2 + g_2^2}
& g^2 \left(\frac{g_1}{g_2} v_1^2 - \frac{g_2}{g_1}
v_2^2 \right) \\
g^2  \left(\frac{g_1}{g_2} v_1^2 - \frac{g_2}{g_1}
v_2^2 \right)
&
g^2 (v_1^2 + v_2^2)
\end{array}
\right)~.
\eeq

The mass eigenvalues for light $W_l$ and heavy $W_h$ bosons can be easily obtained by diagonalizing the above mass matrix. For convenience, we give the approximate expression to order $\epsilon^2 = v^2/u^2$ as follows:
\bea
\label{eq:mwl}
m_{W_l}^2 &=& \frac{1}{2} g^2 v^2 - \frac{1}{2} g^2 v^2 (\sbet^2 -
\sE^2)^2 \epsilon^2 + O(\epsilon^4)~,\nn\\
\label{eq:mwh}
m_{W_h}^2 &=& \frac{1}{2} g^2 u^2 \frac{1}{s_E^2 c_E^2} [ 1 + (\sbet^2
- 2 \sbet^2 \sE^2 + \sE^4) \epsilon^2 ] + O(\epsilon^4)~.
\eea
The lighter $W_l$ boson corresponds to the SM $W$ boson, and has almost the same mass as that in the SM, except for a correction of order $\epsilon^2$. The heavier $W_h$ has a squared mass around $(1/2) g^2 u^2$.  The $W_L$ and $W_H$ fields are almost the mass eigenstates.  The mixing angle $\omega$ defined by
\begin{eqnarray}
W_l = c_\omega W_L - s_\omega W_H\;,\;\;W_h = s_\omega W_L + c_\omega W_H\;,
\end{eqnarray}
is given, to order $\epsilon^2$, by
\beq
\tan 2 \omega = 2 \sE \cE (\cE^2 \sbet^2 - \cbet^2 \sE^2) \epsilon^2 +
O(\epsilon^4)~.
\eeq
Since in our Higgs sector, we anticipate a large $\tan\beta$, to a good approximation we can set $s_\beta^2$ to unity and $c_\beta^2$ to zero.  The charged currents of the quarks are
\bea
{\mathcal L}_W &=& {g_1\over \sqrt{2}} W_1^\mu [ \bar{u} \gamma_\mu P_L d + \bar{c}
\gamma_\mu P_L s ] + {g_2\over \sqrt{2}} W_2^\mu \bar{t} \gamma_\mu P_L b \nn\\
& = & {g\over \sqrt{2}}\, W_L^\mu [ \bar{u} \gamma_\mu P_L d + \bar{c}
\gamma_\mu P_L s + \bar{t} \gamma_\mu P_L b ]  \\
&& + {g\over \sqrt{2}}\, W_H^\mu \left[ \frac{s_E}{c_E} ( \bar{u} \gamma_\mu P_L d + \bar{c}
\gamma_\mu P_L s ) - \frac{c_E}{s_E} \bar{t} \gamma_\mu P_L b \right]\;.\nn
\eea
In the quark mass eigenstate basis, we have
\bea
{\mathcal L}_W
&\approx& {g\over \sqrt{2}} W_l^\mu [\bar{U}_{L} \gamma_\mu V_{KM} D_{L} -
\omega \bar{U}_{L} \gamma_\mu T_U^\dagger N T_D D_{L}] \nn \\
&& + {g\over \sqrt{2}} W_h^\mu [\bar{U}_{L} \gamma_\mu T^\dagger_U N T_D D_{L} +
\omega \bar{U}_{L} \gamma_\mu V_{KM} D_{L}]~,
\label{eq:quarkCC}
\eea
where
\beq
N \equiv {\rm diag}\left(\frac{s_E}{c_E}, \frac{s_E}{c_E},
-\frac{c_E}{s_E}\right) = {\rm diag}\left(\frac{g_1}{g_2}, \frac{g_1}{g_2},
-\frac{g_2}{g_1}\right)~,
\eeq
and $P_L$ is the projection operator for the left-handed currents.
%, and the charged current for leptons are similar.
Hence, $W_L$ has the same coupling as the SM $W$ boson, but has a small mixing with the heavier $W_H$.  On the other hand, $W_H$ couples differently to the third family compared to the first two, depending on the values of $g_1$ and $g_2$.  In Eq.~(\ref{eq:quarkCC}), we have taken the approximations $\sin\omega \approx \omega$ and $\cos\omega \approx 1$ for small mixing angle $\omega$ and kept only terms up to order $\epsilon^2$.

Similarly, we can obtain the charged currents for leptons by replacing $U_L$ and $D_L$ with $\nu_L$ and $E_L$, respectively.  Since the couplings involving the charged leptons in the first two generations are different than that for the third generation, the universality of leptonic charged currents is affected and can result in observable effects.  We will consider the universality of the charged current interactions later.

The mass-squared matrix for the neutral gauge bosons in the basis of the third components $Z_{1,2}$ of the $SU(2)_{1,2}$ gauge bosons and the $U(1)_Y$ gauge boson $B$ is
\beq
M_Z^2 = \frac{1}{2} \left(
\begin{array}{ccc}
g_1^2 (v_1^2 + u^2)     & - g_1 g_2 u^2      & -g' g_1 v_1^2 \\
-g_1 g_2 u^2  & g_2^2(v_2^2 + u^2) & -g' g_2 v_2^2 \\
-g' g_1 v_1^2 & -g' g_2 v_2^2      & g' v^2
\end{array}
\right)~,
\eeq
$g'$ is related to $g$ and $e$ by $1/e^2 = 1/g^2 + 1/{g'^2}$, and $e$ is the usual electromagnetic coupling.  The electroweak mixing angle connects these couplings, \ie, $g = e/s_W$ and $g' = e/c_W$ .  It can be easily checked that the photon field $A$ having zero mass is
\begin{eqnarray}
A = {g^\prime g_2 Z_1 + g^\prime g_1 Z_2 + g_1 g2 B\over \sqrt{g^{\prime 2}(g^2_1+g^2_2) + g^2_1 g^2_2}} ~.
\end{eqnarray}

Again it is convenient to work in the basis $(Z_H, Z_L, A)$.  In the limit of $v_i$ going to zero, the mass of $Z_L$, corresponding to the SM $Z$ boson, also goes to zero.  We find
\begin{eqnarray}
\left ( \begin{array}{c} Z_1\\Z_2\\B
\end{array} \right )
= \left ( \begin{array}{ccc}
g_1/n_1&g_1 g^2_2/n_2&g^\prime g_2/n_3\\
-g_2/n_1&g_2g_1^2/n_2&g^\prime g_1/n_3\\
0&-g^\prime (g^2_1+g^2_2)/n_2&g_1 g_2/n_3\end{array}\right )
\left ( \begin{array}{c} Z_H\\Z_L\\A
\end{array} \right )\;,
\end{eqnarray}
where $n_1 = \sqrt{g^2_1+g^2_2}$, $n_2 = \sqrt{[g^2_1g^2_2+g^{\prime 2} (g^2_1+g^2_2)] (g^2_1+g^2_2)}$ and $n_3 = \sqrt{g^2_1g^2_2+g^{\prime 2}(g^2_1+g^2_2)}$.

In the new $(Z_H, Z_L, A)$ basis, we have
\beq
M_Z^2 = \frac{1}{2} \left(
\begin{array}{ccc}
 (g_1^2 + g_2^2) u^2 + \frac{g_1^4 v_1^2 + g_2^4 v_2^2}{g_1^2 + g_2^2}
& g \sqrt{g^2 + g'^2} \left(\frac{g_1}{g_2} v_1^2 - \frac{g_2}{g_1}
v_2^2 \right)
& 0 \\
g \sqrt{g^2 + g'^2} \left(\frac{g_1}{g_2} v_1^2 - \frac{g_2}{g_1}
v_2^2 \right)
& (g^2 + g'^2) (v_1^2 + v_2^2)
& 0 \\
0 & 0 & 0
\end{array}
\right)~.
\eeq
Because the off-diagonal terms are non-zero, the $Z_H$ and $Z_L$ fields are not mass eigenstates.  The squared masses of the lighter and heavier $Z$ bosons, $Z_l$ and $Z_h$, are
\bea
m_{Z_l}^2 &=& \frac{1}{2} g^2 v^2 \frac{1}{c_W^2} - \frac{1}{2} g^2 v^2
\frac{1}{c_W^2} (\sbet^2 - \sE^2)^2 \epsilon^2 + O(\epsilon^4)~, \nn\\
m_{Z_h}^2 &=& \frac{1}{2} g^2 u^2 \frac{1}{\sE^2  \cE^2} + \frac{1}{2}
g^2 u^2 \frac{(\sbet^2- 2 \sbet^2 \sE^2 + \sE^4 )}{\sE^2  \cE^2} \epsilon^2 +
O(\epsilon^4)~.
\eea
The light $Z_l$ boson reproduces the SM $Z$ boson mass, except for a correction of order $\epsilon^2$.  The mixing angle between $Z_L$ and $Z_H$ is
\beq
\tan 2 \zeta = \frac{2 s_E c_E}{c_W}(\cE^2
\sbet^2 - \sE^2 \cbet^2 ) \epsilon^2 + O(\epsilon^4)~,
\eeq
Note that to order $\epsilon^2$, $W_h$ and $Z_h$ are degenerate.  This is an important test of this model.

In this basis the neutral current interactions can be written as
\bea
{\mathcal L}_{\rm neutral} &=& \bar{\psi} \gamma_\mu \left\{ \frac{1}{2} g' B^\mu
Y + g_1 Z_1^\mu
T_3^1 + g_2 Z_2^\mu T_3^2 \right\} \psi
\nn \\
&=& \bar{\psi} \gamma_\mu \left\{ A^\mu Q + \frac{g}{c_W} Z_L^\mu [
(T_3^1 + T_3^2) - s_W^2 Q] + g Z_H^\mu
\left[ \frac{\sE}{\cE} T_3^1 - \frac{\cE}{\sE} T_3^2 \right] \right\}
\psi \nn \\
&\approx& \bar{\psi} \gamma_\mu \left\{ A^\mu Q + g_Z Z_l^\mu \left[
T_3 - s_W^2 Q - \epsilon^2 (s_E^2 c_E^2 T_3^1
- c_E^4 T_3^2) \right] \right.  \\
&& \left. + g Z_h^\mu \left[ \frac{s_E}{c_E} T_3^1 - \frac{c_E}{s_E}
T_3^2 + \epsilon^2 \frac{s_E c_E^3}{c_W^2} (T_3 - s_W^2 Q)
\right] \right\}\nn
\psi~,
\eea
where $\psi$ can be one of the left- or right-handed quarks and leptons, $Q = Y/2 + T_3^1 + T_3^2$ with $T_3^1$ and $T_3^2$ being the isospin generators for $SU(2)_1$ and $SU(2)_2$, respectively, and $g_Z = g/c_W$.  Since both $SU(2)$ groups are left-handed, $T_3^1$ and $T_3^2$ are both non-zero for left-handed fields only.  Moreover, $T_3^1$ is zero for the third family and $T_3^2$ is zero for the first two.  Here we have assumed small mixing angle $\zeta$ and large $\tan\beta$.

One can easily translate the above interactions to those in the quark mass eigenstates.  There are FCNC interactions due to exchanges of $Z_{l,h}$ at the tree level. They are given by
\begin{eqnarray}
{\mathcal L}_{\rm FCNC} = ({g_Z\over 2} c^2_E \epsilon^2 Z^\mu_l -  {g\over 2 c_E s_E}Z^\mu_h) (\bar U_L \gamma_\mu T^\dagger_U \Delta T_U U_L - \bar D_L \gamma_\mu T^\dagger_D \Delta T_D D_L) ~,
\end{eqnarray}
where $\Delta$ is a diagonal matrix given by $\Delta = {\rm diag}(0,0,1)$. The $Z_l$ FCNC coupling is a special case discussed in Ref.\cite{german}.

\section{Comparing with the SM
\label{sec:pheno}}

\subsection{Precision Test of the Model}

In comparison with the SM, we require $e^{\rmSM} = e$, $G_F^{\rmSM} = G_F$, and $m_Z^{\rmSM} = m_{Z_l}$.   Hereafter, we denote all SM parameters with a subscript 0, {\it e.g.}, $x_0 = \sin^2\theta_W^{\rmSM}$.  Our input parameters are the observed values of $e$, $G_F$ and $m_{Z_l}$ in the new model as they are in the SM.  An important point to remember is that the value of $G_F$ comes from the $\mu$ decay.  We now have two $W$'s contributing to this process: $W_l$ and $W_h$, and  the mixing parameter in $W_l$ also has to be retained.  We get the following relations between the new VEV $v$ ,coupling $g$ and $x = \sin\theta_W$ and the SM parameters,
\bea
\label{eq:vev}
v &=& v_0 [ 1 + \frac{1}{2} \epsilon^2 (1-2c_E^2)^2]\,, \nonumber\\
x &=& x_0 \left[ 1 + \frac{1-x_0}{1-2 x_0}  f_E
\epsilon^2 \right]\,, \\
g &=& g_0 \left[ 1 - \frac{1}{2} \frac{1 - x_0}{1 - 2 x_0}f_E
\epsilon^2 \right]\,.\nonumber
\eea
Hence,
\beq
\label{eq:gZ}
g_Z = \frac{g}{c_W} = {g_Z}_0 \left[ 1 - \frac{1}{2} f_E
\epsilon^2 \right]\,.
\eeq
Here we define $f_E = 1 - 4 \cE^2 + 3 \cE^4$.  The vector and axial-vector couplings of $Z_l$ to fermions are summarized in Table~\ref{tbl:gva}.

\begin{table}[htb]
\begin{center}
\begin{tabular}{|c|c|c|}
\hline
Fermions & $g_V/{g_Z}$ & $g_A/{g_Z}$ \\
\hline
%% nu_e nu_mu
$\nu_e$, $\nu_\mu$  &  $\frac{1}{4}(1 -  \cE^2\sE^2
\esq)$  &  $-\frac{1}{4}(1 -  \cE^2\sE^2 \esq)$ \\ \hline
%% nu_tau
$\nu_\tau$  &  $\frac{1}{4}(1 +  \cE^4  \esq)$
&  $-\frac{1}{4}(1 + \cE^4  \esq$) \\ \hline
%% e mu
$e$, $\mu$  &  $ \frac{1}{4}(-1 +4 x +\cE^2\sE^2
 \esq)$
&  $\frac{1}{4}(1 -  \cE^2\sE^2 \esq)$ \\ \hline
%% \tau
$\tau$  &  $ \frac{1}{4}(-1 + 4x - \cE^4  \esq)$
&  $\frac{1}{4}(1 + \cE^4 ) \esq)$ \\ \hline
%% u, c
$u, c$  &  $\frac{1}{4}(1 - \frac{8}{3} x - \sE^2\cE^2 \esq)$
&  $ \frac{1}{4}(-1 +  \cE^2\sE^2 \esq)$ \\ \hline
%% d, s
$d, s$  &  $\frac{1}{4} (-1 + \frac{4}{3} x +\cE^2 \sE^2  \esq)$
&  $\frac{1}{4}(1 - \cE^2\sE^2 \esq)$ \\ \hline
%% b
$b$  &  $ \frac{1}{4}(-1 + \frac{4}{3} x - \cE^4 \esq)$
&  $\frac{1}{4}(1 + \cE^4 ) \esq)$ \\
\hline
\end{tabular}
\end{center}
\caption{Couplings of the $Z_l$ boson to fermions, in units of the corresponding SM coupling $g_Z$.}
\label{tbl:gva}
\end{table}

The $\rho$ parameter is now
\beq
\rho = \frac{(g^2 + g'^2) m_{W_l}^2}{g^2 m_{Z_l}^2} =
1 - \frac{s_W^2 \cE^2 \sE^2 (\sbet^2 - \sE^2)^2}{c_W^2} \epsilon^4 +
O(\epsilon^6)~.
\eeq
It is interesting to note that the correction is of ${\cal O}(\epsilon^4)$.

As mentioned before, we assume that the measured $m_Z$ is $m_{Z_l}$ in our model.  We now consider a whole range of parameters measured at the $Z$ pole that are used in precision tests of the SM.  We consider shifts from loop-corrected SM predictions of all these parameters to order $\epsilon^2$.  We express all observables in terms of the SM expressions of $x_0, g_0, {g_Z}_0$ through Eqs.~(\ref{eq:vev}) and (\ref{eq:gZ}):
\bea
&& \Gamma_Z = \Gamma_Z^{\rm SM} [ 1 + (-1.35+ 3.70 \cE^2 -1.8 \cE^4) \esq ]~,
\nonumber \\
&& R_{\rm e} = R_{\rm e}^\rmSM [ 1 + (-0.28+1.41 \cE^2 - 0.63 \cE^4) \esq ]~,
\nonumber \\
&& R_{\tau} = R_{\tau}^\rmSM [ 1 + (-0.28 -0.73 \cE^2 - 0.63  \cE^4) \esq ]~,
\nonumber \\
&& R_{b} = R_{b}^\rmSM [ 1 + (0.06 +1.59 \cE^2 + 0.14 \cE^4) \esq ]~,
\nonumber \\
&& R_{c} = R_{c}^\rmSM [ 1 + (-0.12 - 0.12 \cE^2 - 0.27 \cE^4) \esq ]~,
\nonumber \\
&& A_{e,\mu} = A_{e,\mu}^\rmSM [ 1 +  (-17.4 + 57.4 \cE^2 - 40 \cE^4 ) \esq ]~,
\nonumber \\
&& A_{\tau} = A_{\tau}^\rmSM [ 1 + (-17.4+ 69.6 \cE^2 - 40 \cE^4 )\esq ]~,
\nonumber \\
&& A_{u,c} = A_{u,c}^\rmSM [ 1 +  (-1.7 +5.64 \cE^2 - 3.9  \cE^4) \esq ]~,
 \\
&& A_{d,s} = A_{d,s}^\rmSM [ 1 + (-0.22 +0.74 \cE^2 -0.52  \cE^4)\esq ]~,
\nonumber \\
&& A_{b} = A_{b}^\rmSM [ 1 +  (-0.22 +0.90 \cE^2 -0.52  \cE^4) \esq]~,
\nonumber \\
&& A_{FB}^e = {A_{FB}^e}^\rmSM [ 1 + (-34.8 +114.8\cE^2 - 80.0\cE^4)\esq ]~,
\nonumber \\
&& A_{FB}^\tau = {A_{FB}^\tau}^\rmSM [ 1 + (-34.8 +126.9\cE^2 -80.0\cE^4 ) \esq ]~,
\nonumber \\
&& A_{FB}^{u,c} = {A_{FB}^{u,c}}^\rmSM [ 1 + (-19.1 +63.0 \cE^2 -43.9
\cE^4 ) \esq ]~,
\nonumber \\
&& A_{FB}^{d,s} = {A_{FB}^{d,s}}^\rmSM [ 1 + (-17.6 +58.13 \cE^2 -40.5\cE^4)\esq ]~,
\nonumber \\
&& A_{FB}^b = {A_{FB}^b}^\rmSM [ 1 + (-17.6 +58.29 \cE^2 -40.5 \cE^4)\esq ]~,\nonumber
\label{eq:Zpole}
\eea
%
%CC added a footnote below
All SM quantities above include radiative corrections \footnote{We note that EWP corrections in many models with extended groups have been considered in Ref.~\cite{EWP}.  Our results differ from theirs because of different inputs and new data.}.

As mentioned earlier, our model also predicts violation of universality in charged lepton decays.  We now consider the constraint obtained from this consideration.  First, there is no violation of universality for the first two generations in the model. Therefore, the universality between $\tau \to \mu \bar \nu_\mu \nu_\tau$ and $\tau \to e \bar \nu_e \nu_\tau$ are not affected.  But they are different from the $\mu \to e \bar \nu_e \nu_\mu$ process.  We will thus compare $\tau \to (\mu,e) \bar \nu_{\mu,e} \nu_\tau$ with $\mu \to e \bar \nu_e \nu_\mu$.  The decay widths of these modes
\begin{eqnarray}
\Gamma \propto \frac{G_{\ell\ell'}}{192\pi^3} m_\ell^5 ~,
\end{eqnarray}
where $\ell$ and $\ell'$ denote the leptons in the initial and final states, respectively.  As said above, we take $G_{\mu e}$ as the SM $G_F$.  Then the model gives
\begin{eqnarray}
\frac{G_{\tau e}^2}{G_F^2} = \frac{G_{\tau\mu}^2}{G_F^2}
= \left[ 1 - \epsilon^2 ( 1 - 2c_E^2 ) \right]^2 ~.
\end{eqnarray}
Note that the corrections here are also of order $\epsilon^2$ at the amplitude level.  Experimentally\cite{PDG2009},
\begin{eqnarray}
\frac{G_{\tau e}^2}{G_F^2} &=& 1.0012 \pm 0.0053 ~, \nonumber \\
\frac{G_{\tau\mu}^2}{G_F^2} &=& 1.0087 \pm 0.0185 ~,
\label{eq:lepuniv}
\end{eqnarray}
respectively.  Here we have taken into account the finite $m_\mu$ phase space effect in the second line of Eq.~(\ref{eq:lepuniv}).

We now combine the above-mentioned EWP data, Eq.~(\ref{eq:Zpole}), and the lepton universality constraints, Eq.~(\ref{eq:lepuniv}), to perform a global fit to available data \cite{PDG2009} for our theory parameters, $c_E$ and $\epsilon$.  The best-fitted values are $\cE = 0.633$ and $\epsilon = 0.059$ with $\chi^2_{\rm min} = 16.28$, in comparison with the SM $\chi^2_{\rm min} = 18.86$.  These values of parameters correspond to both $m_{W_h}$ and $m_{Z_h}$ around $2.8$~TeV, well within the reach of the LHC.

Since $m_{Z_l}$ is fixed to the experimentally measured value $m_Z^{\rm SM}$ in our analysis, the value of $m_{W_l}$ is shifted from the SM value in the following way
\begin{eqnarray}
m_{W_l}^2 - {m_W^{\rm SM}}^2 =
-{m_W^{\rm SM}}^2 \frac{f_E \epsilon^2 x_0}{1 - 2x_0} ~.
\end{eqnarray}
Therefore, $m_{W_l}$ is smaller than $m_Wx^{\rm SM}$ by about $7$ MeV.  This is well within the uncertainties after taking into account radiative corrections due to Higgs and top quark exchanges.  We have also verified the effect of our modification on atomic parity violation experiments. The change in value of $Q_W$ is 0.1 \% and is too small to be observed.

\subsection{FCNC in the Model}

In this model there are two types of tree-level FCNC's, with one from $Z_l$ and $Z_h$ exchanges and the other from Higgs exchanges.  The relevant parts are given by
\begin{eqnarray}
&&{\mathcal L}_{Z-\rm FCNC} =
\left( {g_Z\over 2} c^2_E \epsilon^2 Z^\mu_l -  {g\over 2 c_E s_E}Z^\mu_h \right) \bar f_L \gamma_\mu \tilde \Delta^z_f T_3 f_L\;,\nonumber\\
&&{\mathcal L}_{Y-\rm FCNC} =
\left( 1+{c_\beta\over s_\beta} \right)\left [\bar U_R \tilde \Delta^Y_u U_L (H^0 - iA^0)
 + \bar D_R \tilde \Delta^Y_d D_L (H^0 + i A^0)\right ] ~,
\end{eqnarray}
where $\tilde \Delta^z_f = T^\dagger_f \Delta T_f$ and $\tilde \Delta^Y_f =  S_f \lambda^f_1 T^\dagger_f$.

Since the interactions depend on the unknown mixing matrices $S_i$, $T_i$ and $\lambda^i_1$ even if we know the mass scale of new physics, it is not possible to
make definite predictions.  There are many FCNC processes which can be used to constrain the parameters.  A complete FCNC analysis is out of the scope of this paper.  We will, as an example, show that the central values of $\epsilon$ and $c_E$ are allowed by the FCNC constraint from recent $B_{d,s}$-$\bar B_{d,s}$ mixing data.

At the quark level, the contributions to the mixing from the above gauge and Yukawa interactions are give by
\begin{eqnarray}
M_{12} &=&
\left [{g^2_Z\over 4 m^2_{Z_l}}( c^2_E\epsilon^2)^2 + {g^2\over 4 c^2_Es^2_E m^2_{Z_h}}\right ]
\langle B_{q}|(\tilde \Delta^z_{qb}\bar q \gamma^\mu L b )^2|\bar B_b \rangle \nonumber\\
&+& {1\over m^2_{H}} \left( 1+{c_\beta\over s_\beta} \right)^2
\langle B_q | \left (\bar q (\tilde \Delta^Y_{qb}L  + \tilde \Delta^{Y*}_{bq} R) b \right )^2|\bar B_q \rangle \\
&-&{1\over m^2_{A}} \left( 1+{c_\beta\over s_\beta} \right)^2
\langle B_q | \left (\bar q (\tilde \Delta^Y_{qb}L  - \tilde \Delta^{Y*}_{bq} R) b \right )^2|\bar B_q \rangle ~.\nn
\end{eqnarray}
For the gauge interaction, the contribution from $Z_l$ exchange is suppressed by $\epsilon^4$ and can be neglected compared with that from $Z_h$ exchange. Using the leading approximation $m^2_{Z_l}/m^2_{Z_h} = \epsilon^2c^2_Es^2_E/c^2_W$, we have a simple expression
\begin{eqnarray}
M_{12} = {g_Z^2\over 4 m^2_{Z_l}}\epsilon^2
\langle B_{q}|(\tilde \Delta^z_{qb}\bar q \gamma^\mu L b )^2|\bar B_b \rangle ~.
\end{eqnarray}

The effective coupling characterizing the contribution to the mixing is $\epsilon \tilde \Delta_{qb}$.  In general, they are not known and can be constrained from available data. If it turns out that the couplings are given by the two scenarios in Section~\ref{sec:yukawa}, we will obtain for Case a):
\begin{eqnarray}
\mbox{Case a)}:&&\tilde \Delta_{db} = \epsilon V^*_{td} V_{tb} \approx  5 \times 10^{-4}\;,\;\;\tilde \Delta_{sb} = \epsilon  V^*_{ts} V_{tb} \approx 2.5 \times 10^{-3} ~.\nonumber
\end{eqnarray}
Taking the above couplings as an estimate, we find that these are one order of magnitude smaller than the experimental bounds on these couplings.

For the Higgs exchange contributions with Case b), we have $\tilde \Delta^Y_d = \lambda_1^d V_{KM}$.  As long as $\tilde \Delta^Y_{db, bd}$ and $\tilde \Delta^Y_{sb, sd}$ are not too much larger than $5\times 10^{-3}$ and $2.5\times 10^{-2}$, the heavy Higgs masses can be as low as $2.7$ TeV, as allowed for $m_{Z_h}$.

We have also checked constraints on gauge boson exchanges that come from rare $B$ decays and $K \bar K$ mixing. These contributions are highly suppressed with the allowed values of $\epsilon$ and $c_E$ and offer no constraints.  There are also FCNC interactions involving charged leptons. These interactions are determined by another set of parameters similar to what we have discussed for the quark sector. Since these parameters are in principle independent of the parameters in the quark sector, one can always adjust the parameters to satisfy experimental bounds without spoiling the relatively low mass new gauge bosons allowed by the precision tests discussed earlier.

We conclude that the FCNC parameters can be easily adjusted to be consistent with data while allowing the heavy gauge boson and Higgs boson masses to be as low as a few TeV.

\section{Summary
\label{sec:summary}}

Motivated by the reach of the LHC for discovery of heavy gauge bosons, we have explored the family $SU(2)_l \times SU(2)_h \times U(1)$ model.  Such a model can throw some light on the origin of the family structure. We confront the model with electroweak precision data on one hand and consistency in the Higgs sector on the other.  We conclude from the best fit, which has a slightly lower $\chi^2_{\rm min}$ than the SM, that the best values for the model parameters are $\cE = 0.633$ and $\epsilon = 0.059$.  This yields for the heavy gauge boson masses:
\beq
m_{W_h} = m_{Z_h} = m_{W_l}/(\sE \cE \epsilon)= 2.77 {\rm TeV}
\eeq
This value is substantially higher than what previous studies have assumed. Besides, in consideration of FCNC effects, we find that the heavy Higgs doublet is also at least as high in mass.  The gauge sector in the model also exhibits characteristic violation of universality, which distinguishes this class of models from others that have large-mass gauge bosons.

\begin{acknowledgments}

  This research was supported in part by the U.S.~Department of Energy under Grant No.\ DE-FG02-96ER40969 and in part by the National Science Council of Taiwan, R.O.C.\ under Grant No.\ NSC 94-2112-M-008-023- and No.~NSC~97-2112-M-008-002-MY3.  C.-W.~C. and X.-G.~H. would like to thank the partial support of National Center for Theoretical Sciences, Taiwan.

\end{acknowledgments}

\end{document}